\documentclass[twocolumn,aps,showpacs,floatfix,superscriptaddress]{revtex4}
\pdfoutput=1 
\usepackage{amsmath,amssymb,eucal,graphicx}
\usepackage[all]{xy}

\begin{document}
\title{Outbreak size distributions in epidemics with multiple stages}

\author{Tibor Antal}
\affiliation{School of Mathematics, Edinburgh University, Edinburgh, EH9 3JZ, UK}
\author{P. L. Krapivsky}
\affiliation{Department of Physics, Boston University, Boston, MA 02215, USA}

\begin{abstract}
Multiple-type branching processes that model the spread of infectious diseases are investigated. In these stochastic processes, the disease goes through multiple stages before it eventually disappears. We mostly focus on the critical multistage Susceptible-Infected-Recovered (SIR) infection process. In the infinite population limit, we compute the outbreak size distributions and show that asymptotic results apply to more general multiple-type critical branching processes. Finally using heuristic arguments and simulations we establish scaling laws for a multistage SIR model in a finite population.  
\end{abstract}
\maketitle

\section{Introduction}

A critical branching process (CBP) is the fundamental stochastic process that has innumerable applications \cite{feller,teh,athreya04,branch,vatutin}. Although the adjective ``critical'' suggests that the CBP is a peculiar branching process which should require the tuning of parameters, the CBP is arguably more important than its generic sub-critical and super-critical brethren. Indeed, the sub-critical branching process quickly leads to extinction while the super-critical branching process may result in an unlimited growth, so whenever we are seeking a framework for phenomena which are driven by branching and are maintained in a stationary state, the CBP provides the proper setting. 

The simplest branching process involves duplication and death, and hence called a birth-death process. Symbolically 
\begin{displaymath}
\xymatrix{A+A                & \\
               A \ar[u] \ar[r]   & \emptyset}
\end{displaymath}
For the critical birth-death process the probabilities of these two events are equal. One can think on $A$s  being cells which either divide or die. Another interpretation is to identify $A$s with infected individuals who either die or transmit infection to other individuals. The branching process then becomes a stochastic version of the Susceptible-Infected-Recovered (SIR) infection process. According to this model \cite{ntjb,ma,hwh,jdm}, the population consists of susceptible, infected, and recovered individuals, and the infection spreads by contact between infected and susceptible members of the community, while infected individuals recover (and become immune), or equivalently they are removed from the system. The branching process also provides the simplest model of biodiversity \cite{Aldous} (see \cite{Nee} for a survey of biological literature on macroevolution modeling), with duplication and death processes corresponding to speciations and extinctions. The CBP is {\em neutral} as it assumes that speciations and extinctions occur with equal probability. In the context of infectious diseases the CBPs naturally arise as a balance between human efforts leading to the reduction of the infection rate \cite{ma} and the natural evolution that increases the infection rate of diseases hovering below the threshold \cite{antia}. 

In numerous applications, it is desirable to generalize the CBP to include more than one type of elemental building blocks. In cell biology, we may have two populations of cells, progenitor $A$ cells that can divide and differentiate into $B$ cells. These two cell populations can be (on average) in a stationary state (homeostasis), if they evolve according to the two-type CBP. The simplest two-type CBP is represented by the scheme
\begin{displaymath}
\xymatrix{A+A                 & B+B               & \\
               A \ar[u] \ar[r]    & B\ar[u] \ar[r]   & \emptyset}
\end{displaymath}
where all steps are equiprobable. 

Branching processes can be studied in discrete time when all infected individuals divide or die synchronously, or in continuous time when each infected individual independently divides or dies. For a critical process the birth and death probabilities or rates are equal. Continuous-time two-type birth-death branching processes are solvable, namely the time-dependent sizes of populations of $A$ and $B$ individuals have been analytically determined \cite{skin&cancer}. Specializing the results of Refs.~\cite{skin&cancer} to the above simplest two-type critical CBP leads to simplifications, yet the results are still cumbersome. In this paper we are concerned with different questions, namely we are interested in time-independent quantities summarizing the outbreak size distributions. 

The rest of this paper is organized as follows.  In Sec.~\ref{CBDP} we consider the simplest multiple-type critical branching process and derive, in the infinite population limit, the outbreak size distributions. In Sec.~\ref{more} we show the universality of the asymptotic results for the outbreak size distributions, namely that at the epidemic threshold these distributions have algebraic tails which are determined only by the number of types (but not the detail of the model). In Sec.~\ref{finite} we analyze finite populations and obtain finite-size scaling laws for the sizes of the outbreaks. In Sec.~\ref{summary} we give a short discussion. Finally, in an appendix, some of the details of the analysis of a generating function are presented. 

\section{Critical birth-death process}
\label{CBDP}

In this section we consider the infinite population limit. We shall always assume that the system begins with a single $A$, representing an infected individual. In the multi-type case, an $A$ can turn into $B$, a $B$ into $C$, and so on. The probability for each individual to duplicate or disappear is the same, which makes our process critical. Consequently, all individuals eventually disappear with probability one and the epidemic stops. We want to determine the outbreak size distribution.

\subsection{Single Type Critical Birth-Death Process} 

Let $A_n$ be the probability that exactly $n$ individuals catch the infection before the epidemic is over. We have $A_1=\tfrac{1}{2}$. Further, $A_2=\tfrac{1}{2}A_1^2$ since at the first step a new individual must get infected and then both must die without spreading infection. Proceeding along these lines we arrive at the recurrence 
\begin{equation}
\label{An_rec}
A_n = \frac{1}{2}\sum_{i+j=n}A_iA_j+\frac{1}{2}\,\delta_{n,1}
\end{equation}
reflecting that the first infection event results in two independent infection processes \cite{teh}.  A solution to \eqref{An_rec} is found by introducing the generating function
\begin{equation}
\label{An_gen}
\mathcal{A}(z) = \sum_{n\geq 1}A_n z^n
\end{equation}
which converts \eqref{An_gen} into a quadratic equation $2\mathcal{A}=\mathcal{A}^2 +z$
whose solution reads
\begin{equation}
\label{Az}
\mathcal{A}(z)  = 1 - \sqrt{1-z}
\end{equation}
Expanding $\mathcal{A}(z)$ in powers of $z$ we find
\begin{equation}
\label{An_sol}
A_n = \frac{1}{\sqrt{4\pi}}\, \frac{\Gamma\left(n-\tfrac{1}{2}\right)}{\Gamma(n+1)}
\sim \frac{1}{\sqrt{4\pi}}\,n^{-3/2}
\end{equation}
In particular, the probabilities $A_n$ are given by 
\begin{equation*}
 \frac{1}{2},\frac{1}{8},\frac{1}{16},\frac{5}{128},\frac{7}{256},\frac{21}{1024},\frac{33}{2048},\frac{429}{32768},\frac{715}{65536},\frac{2431}{262144}
\end{equation*}
for $n=1,\dots,10$. 
Note that the above process can also be interpreted as a discrete time symmetric random walk on the integers, where $A_n$ corresponds to the probability that the first return to the origin occurs at time $2n$ \cite{feller}.

\subsection{Two-Type Critical Birth-Death Process} 

In this case the size distribution of the outbreaks of $A$ types is the same as for the single type CBP. Consider now the outbreaks of $B$ types. Denote by $B_n$ the probability that exactly $n$ individuals of type $B$ are born before the outbreak is over. These probabilities satisfy 
\begin{equation}
\label{Bn_rec}
B_n = \frac{1}{2}\sum_{i+j=n}B_iB_j+\frac{1}{2}\,A_n
\end{equation}
To understand this recurrence it suffices to consider what happens to the initial $A$ individual. One possibility is that the initial $A$ turns into a $B$ (this happens with probability 1/2), and then this $B$ produces $n$ individuals of type $B$ with probability $A_n$. Combining we get the second term $\frac{1}{2}A_n$ on the right-hand side of Eq.~\eqref{Bn_rec}. Otherwise the initial $A$ produces two $A$s (this happens with probability 1/2), and then the probability that these two $A$s will produce exactly $n$ individuals of type $B$ is just the convolution $\sum_{i+j=n} B_i B_j$. Overall this gives the first term on the right-hand side of Eq.~\eqref{Bn_rec}. 

To tackle Eq.~\eqref{Bn_rec} we again proceed by introducing the generating function
\begin{equation}
\label{Bn_gen}
\mathcal{B}(z) = \sum_{n\geq 1}B_n z^n
\end{equation}
which converts \eqref{Bn_gen} into a quadratic equation $2\mathcal{B} = \mathcal{B}^2 +\mathcal{A}$
whose solution reads
\begin{equation}
\label{Bz}
\mathcal{B}(z)  = 1 - (1-z)^{1/4}
\end{equation}
Expanding $\mathcal{B}(z)$ in powers of $z$ we find
\begin{equation}
\label{Bn_sol}
B_n = \frac{1}{4\,\Gamma\left(\tfrac{3}{4}\right)}\, \frac{\Gamma\left(n-\tfrac{1}{4}\right)}{\Gamma(n+1)}
\sim \frac{1}{4\,\Gamma\left(\tfrac{3}{4}\right)}\,n^{-5/4}
\end{equation}

\subsection{Multiple-Type Critical Birth-Death Process} 

Generally for multiple-type CBP we seemingly have
\begin{equation}
\label{Anp_rec}
A_n^{(p)} = \frac{1}{2}\sum_{i+j=n}A_i^{(p)}A_j^{(p)}+\frac{1}{2}\,A_n^{(p-1)}
\end{equation}
The corresponding generating functions $\mathcal{A}^{(p-1)}$ and  $\mathcal{A}^{(p)}$ are related via
\begin{equation}
\label{Ap_gen}
2\mathcal{A}^{(p)} = \big[\mathcal{A}^{(p)}\big]^2+ \mathcal{A}^{(p-1)}
\end{equation}
We already know that
\begin{eqnarray*}
\mathcal{A}^{(1)} &\equiv & \mathcal{A} = 1 - (1-z)^{1/2}\\
\mathcal{A}^{(2)} &\equiv & \mathcal{B} = 1 - (1-z)^{1/4}
\end{eqnarray*}
A general solution to \eqref{Ap_gen} is
\begin{equation}
\label{Apz}
\mathcal{A}^{(p)} = 1 - (1-z)^{1/2^p}
\end{equation}
from which
\begin{equation}
\label{Anp_sol}
A_n^{(p)} = \frac{1}{2^p\,\Gamma\left(1-2^{-p}\right)}\, 
\frac{\Gamma\left(n-2^{-p}\right)}{\Gamma(n+1)}
\sim \frac{n^{-1-2^{-p}}}{2^p \Gamma(1-2^{-p})}
\end{equation}
One can verify \eqref{Apz} by induction. It also follows from the more general result of the following subsection~\ref{joint}. 

\subsection{Joint Outbreak Size Distribution}
\label{joint}

The joint outbreak size distribution first becomes relevant for the two-type CBP. In this setting we want to compute the probability $P_{m,n}$ that exactly $m$ individuals of type $A$ and $n$ individuals of type $B$ are born before the outbreak is over.  

The disappearance of an $A$ results in the birth of a $B$, and therefore $P_{m,n}\equiv 0$ whenever $m>n$. To compute $P_{m,n}=0$ in the interesting $m\leq n$ range, we shall use the important {\em abelian} property of the model: The composition of the system at the moments of birth of new $B$s is irrelevant, so one can assume that the process starts with $m$ individuals of type $B$ (and no $A$ individuals). We have $P_{m,m}=A_m (A_1)^m$ where the first term assures that exactly $m$ individuals of type $A$ ever exist and the second term gives the probability that all $m$ individuals of type $B$ eventually disappear without duplication. Further we have
$P_{m,m+1}=A_m m(A_1)^{m-1}A_2$, and generally
\begin{equation}
\label{Pmn_rec}
P_{m,n} = A_m\sum_{i_1+\ldots+i_m=n}A_{i_1}\ldots A_{i_m}
\end{equation}
Using the generating function 
\begin{equation*}
\mathcal{P}(x,y) = \sum_{n\geq m\geq 1}P_{m,n} x^m y^n
\end{equation*}
and \eqref{Pmn_rec} we get
\begin{equation}
\label{gen2type}
\begin{split}
\mathcal{P}(x,y) &= \sum_{m\geq 1}  A_m x^m \big[\mathcal{A}(y)\big]^m\\
 &= \mathcal{A}(x\mathcal{A}(y)) \\
 &= 1-\sqrt{1-x\mathcal{A}(y)}\\
 &= 1-\sqrt{1-x + x\sqrt{1-y}}
\end{split} 
\end{equation}
Specializing $x$ or $y$ to unity we recover previous results \eqref{Az} and \eqref{Bz}:
\begin{equation}
\mathcal{P}(x,y=1) = \mathcal{A}(x), \quad 
\mathcal{P}(x=1,y) = \mathcal{B}(y)
\end{equation}
This essentially provides another derivation of Eq.~\eqref{Bz}. 

From the generating function \eqref{gen2type} we can express the probabilities more explicitly as
\begin{equation}
\label{2typeexp}
 P_{m,n} = \frac{(2m-3)!!}{2^{m+n} n!} s_{n,n-m+1}
\end{equation}
where the integer $s_{n,k}$ for $n\ge k\ge1$ are given by the recursion
\begin{equation}
\label{recurs}
 s_{n,k} = s_{n-1,k} + (n+k-3) s_{n-1,k-1}
\end{equation}
with $s_{1,1}=1$; otherwise we consider $s_{n,k}$ to be zero. The first few values of $s_{n,k}$ are as follows
\begin{equation}
\left(
\begin{array}{cccccc}
 1 & 0 & 0 & 0 & 0 & 0 \\
 1 & 1 & 0 & 0 & 0 & 0 \\
 1 & 3 & 3 & 0 & 0 & 0 \\
 1 & 6 & 15 & 15 & 0 & 0 \\
 1 & 10 & 45 & 105 & 105 & 0 \\
 1 & 15 & 105 & 420 & 945 & 945
\end{array}
\right)
\end{equation}
For example, the above formulas lead to the following probabilities $P_{m,n}$ for $m,n\le 6$
\begin{equation}
\left(
\begin{array}{cccccc}
 \frac{1}{4} & \frac{1}{16} & \frac{1}{32} & \frac{5}{256} & \frac{7}{512} & \frac{21}{2048} \\
 0 & \frac{1}{32} & \frac{1}{64} & \frac{5}{512} & \frac{7}{1024} & \frac{21}{4096} \\
 0 & 0 & \frac{1}{128} & \frac{3}{512} & \frac{9}{2048} & \frac{7}{2048} \\
 0 & 0 & 0 & \frac{5}{2048} & \frac{5}{2048} & \frac{35}{16384} \\
 0 & 0 & 0 & 0 & \frac{7}{8192} & \frac{35}{32768} \\
 0 & 0 & 0 & 0 & 0 & \frac{21}{65536}
\end{array}
\right)
\end{equation}
One can recognize the first few rows of the matrix  $P_{m,n}$. For the first three rows 
\begin{equation}
\begin{split}
P_{1,n} &= \frac{(2n-2)!}{(n-1)!\, n!}\,\frac{1}{2^{2n}}\\
P_{2,n} &=  \frac{(2n-2)!}{(n-1)!\, n!}\,\frac{1}{2^{2n+1}} \qquad n\geq 2\\
P_{3,n} &=  \frac{(2n-4)!}{(n-3)!\, n!}\,\frac{3}{2^{2n+1}} \qquad n\geq 3
\end{split}
\end{equation}
These results are straightforwardly verified by induction. One can also identify the diagonal elements
\begin{equation}
P_{n,n}  =  \frac{(2n-2)!}{(n-1)!\, n!}\,\frac{1}{2^{3n-1}}
\end{equation}
We haven't succeeded, however, in finding an explicit expression for $P_{m,n}$ in the general case.

Consider now the three-type CBP. We again limit ourselves with the simplest three-type CBP which is represented by the scheme
\begin{displaymath}
\xymatrix{A+A                 & B+B            &      C+C        &\\
                A \ar[u] \ar[r] & B\ar[u] \ar[r] & C\ar[u] \ar[r] &\emptyset}
\end{displaymath}
where all events occur with probabilities $\tfrac{1}{2}$. The probability $P_{m,n,\ell}$ that exactly $(m,n,\ell)$ individuals of types $(A,B,C)$ are born before the outbreak is over is given by an obvious generalization of Eq.~\eqref{Pmn_rec}:
\begin{equation*}
P_{m,n,\ell} = A_m\sum_{i_1+\ldots+i_m=n}A_{i_1}\ldots A_{i_m}
\sum_{j_1+\ldots+j_n=\ell}A_{j_1}\ldots A_{j_n}
\end{equation*}
Converting this recurrence into an equation for the generating function 
\begin{equation*}
\mathcal{P}(x,y,z) = \sum_{\ell\geq n\geq m\geq 1}P_{m,n,\ell} x^m y^n z^\ell
\end{equation*}
we find
\begin{eqnarray*}
\mathcal{P}(x,y,z) &=& \mathcal{A}(x \mathcal{A}(y \mathcal{A}(z))) \\
 &=& 1-\sqrt{1-x + x\sqrt{1-y+y\sqrt{1-z}}}
\end{eqnarray*}

Generally for the $p-$type branching process the generating function
\begin{equation*}
\mathcal{P}(z_1,\ldots,z_p) = \sum_{m_p\geq \ldots\geq m_1\geq 1}
P_{m_1,\ldots,m_p} z_1^{m_1} \ldots z_p^{m_p}
\end{equation*}
which encodes the joint distribution $P_{m_1,\ldots,m_p}$ can be expressed through the generating function $\mathcal{A}$ of the  single type process as
\begin{equation}
\mathcal{P}(z_1,\dots ,z_p) = 
\mathcal{A}(z_1 \mathcal{A}(z_2 \mathcal{A}( \dots \mathcal{A}(z_{p-1}\mathcal{A}(z_p))\dots)))
\end{equation}
Specializing $\mathcal{P}(z_1,\ldots,z_p)$ to $z_1=\ldots=z_{p-1}=1, z_p=z$ we must reproduce the generating function $\mathcal{A}^{(p)}(z)$. We actually get an identity
\begin{equation}
\label{Ap_pA}
\mathcal{A}^{(p)}(z) = \mathcal{A}(\mathcal{A}(\ldots \mathcal{A}(\mathcal{A}(z))\ldots))
\end{equation}
Using \eqref{Ap_pA} in conjunction with \eqref{Az} we indeed recover \eqref{Apz}.

\section{More general processes}
\label{more}

Let us first generalize the results of the previous section to non-critical brith-death processes.
For a birth-death branching process we set the recovery rate to unity and denote by $\alpha$ the infection rate. The process is called subcritical for $\alpha<1$, critical  for $\alpha=1$, and supercritical for $\alpha>1$. While for $\alpha\le 1$ the outbreak is finite with probability one; for a supercritical process ($\alpha>1$) the outbreak never stops with probability $1-1/\alpha$. We can still calculate the outbreak size distribution by conditioning on finite outbreaks. Let $A_n$ be the probability that a single individual results in a finite outbreak where exactly $n$ individuals are infected during the epidemic. We can write the following recursion
\begin{equation}
\label{An:sub_rec}
A_n = \frac{\alpha}{1+\alpha}\sum_{i+j=n}A_iA_j + \frac{1}{1+\alpha}\,\delta_{n,1}
\end{equation}
which, for the generating function takes the form 
\begin{equation}
\label{genf_super}
 (1+\alpha) \mathcal{A} = \alpha \mathcal{A}^2+z
\end{equation}
and its solution is given by
\begin{equation}
\label{Az:sub}
\mathcal{A} = \frac{1+\alpha-\sqrt{(1+\alpha)^2-4\alpha z}}{2\alpha}
\end{equation}
The probability of a finite outbreak is of course
\begin{equation}
\mathcal{A}(z=1) = \left\{
\begin{array}{cc}
 1 &  \mathrm{for}~~ \alpha\le 1 \\
 1/\alpha & \mathrm{for}~~  \alpha\ge1  
\end{array}
\right.
\end{equation}
and the average outbreak size in case of a finite outbreak is
\begin{equation}
\label{nav}
\langle n\rangle=\sum nA_n=\mathcal{A}'(z=1)=\frac{1}{|1-\alpha|}
\end{equation}

By expanding $\mathcal{A}$ in powers of $z$, we obtain the probability of a finite outbreak of size $n$
\begin{equation}
 A_n = \frac{1+\alpha}{4\alpha\sqrt\pi} \left[ \frac{4\alpha}{(1+\alpha)^2}\right]^n
  \frac{\Gamma\left(n-\tfrac{1}{2}\right)}{\Gamma(n+1)}
\end{equation}
Asymptotically, for large $n$ we have a power law decay
\begin{equation}
 A_n  \sim \frac{1+\alpha}{4\alpha\sqrt\pi} e^{-n/\xi} n^{-3/2}
\end{equation}
with an exponential cutoff around
\begin{equation}
  \xi = \left[ 2\log(1+\alpha) - \log 4\alpha \right]^{-1}
\end{equation}
The cutoff diverges as $\alpha\to1$, and for the critical process $\alpha=1$ we find a pure power law decay 
\eqref{An_sol}.

More generally, consider a branching process where each individual is replaced by $k$ individuals at rate $\alpha_k$. Equivalently, we can consider a discrete time branching process, where at every time step each individual is independently replaced by $k$ individuals with probability $p_k=\alpha_k/\sum \alpha_n$. In this case the governing equation for the outbreak distribution (which is essentially the backward Kolmogorov equation) is a simple generalization of Eq.~\eqref{genf_super},  namely
\begin{equation}
\label{stgenrec}
\mathcal{A} \sum_{k\ge 0} \alpha_k  = \sum_{k\ge1} \alpha_k \mathcal{A}^k + \alpha_0 z
\end{equation}
An explicit solution becomes less accessible for larger number of possible offspring, but for example the critical process with three offspring ($\alpha_0=2/3$, $\alpha_3=1/3$; other rates are zero) with a single initial individual is still tractable
\begin{equation}
 \mathcal{A} =  2 \sin \frac{\arcsin z}{3} 
\end{equation}
Note that in this case it is impossible to find a cluster with even number of individuals: $A_{2n}\equiv 0$. The first few nonzero values of $A_n$ are
\begin{equation*}
 A_1=\frac{2}{3}, A_3= \frac{8}{81}, A_5= \frac{32}{729}, A_7= \frac{512}{19683}, A_9= \frac{28160}{1594323}
\end{equation*}
From the behavior of $\mathcal{A}(z)$ around $z=1$ we find that for large odd $n$ the asymptotic is 
\begin{equation}
 A_n \sim \frac{1}{\sqrt{6\pi}}\, n^{-3/2}
\end{equation}
The exponent is again equal to $3/2$. Using \eqref{stgenrec} one can show that this exponent is universal for any (single-type) branching process. We state this for the equivalent discrete time version where the probability of a individual having $k$ offspring is $p_k=\alpha_k/\sum_n \alpha_n$ for $k=0,1,2,\dots$:

\medskip\noindent
{\it Consider a discrete time branching process with a single initial individual and an offspring probability function $p_k$. Suppose that the process is critical, $\sum k p_k=1$, and has finite variance $\sigma^2=\sum k^2 p_k - 1<\infty$. Then asymptotically $1-\mathcal{A}\sim \sqrt{c(1-z)}$ for $z\to 1$, and 
\begin{equation}
\label{CBP_1}
A_n\sim \sqrt{\frac{c}{4\pi}}\, n^{-3/2}
\end{equation}
for $n\to\infty$, with $c=2p_0/\sigma^2$.}

\medskip\noindent
To prove this assertion we recall that $\mathcal{A}(z)$ is analytic for $|z|<1$ and $\mathcal{A}(1)=1$. We write $\mathcal{A}(z)=1-\mathcal{B}(z)$, where $\lim_{z\to 1}\mathcal{B}(z)=0$, and plug this expression into \eqref{stgenrec}. Using the conditions for the zeroth and first moments of $p_k$, we find
\begin{eqnarray*}
 1-\mathcal{B} &=& \sum_{k\ge 0} p_k (1-\mathcal{B})^k - p_0(1-z)\\
 & = & 1-\mathcal{B}+\frac{\mathcal{B}^2\sigma^2}{2} + O(\mathcal{B}^3) - p_0(1-z)
\end{eqnarray*}
which gives the leading asymptotic behavior 
\begin{equation}
 \mathcal{B} = \sqrt{c(1-z)}+ O(1-z)
\end{equation}
with $c=2p_0/\sigma^2$. Expanding $\sqrt{1-z}$, or equivalently using a Tauberian theorem \cite{feller}, we arrive at the asymptotic \eqref{CBP_1} for $A_n$, which completes the proof. 

The asymptotic behavior of multi-type processes now follows from recursion \eqref{Ap_pA}:
\begin{equation*}
\begin{split}
1-\mathcal{A}^{(p)}(z) &= 1- \mathcal{A}( \mathcal{A}^{(p-1)}(z) ) \sim c^{1/2} [1-\mathcal{A}^{(p-1)}(z)]^{1/2}\\
&\sim c^{1/2} c^{1/4} [1-\mathcal{A}^{(p-2)}(z)]^{1/4} \sim \ldots\\  
&\sim c^{1-2^{-p}} (1-z)^{2^{-p}}
\end{split}
\end{equation*}
If at all stages there are different critical branching processes, then simply
\begin{equation}
 1-\mathcal{A}^{(p)}(z) \sim c_1^{1/2}  c_2^{1/4} \cdots c_p^{1/2^{p}} (1-z)^{2^{-p}}
\end{equation}
where all $c_j$ are defined analogously. This corresponds to the asymptotic behavior for the probabilities
\begin{equation}
 A_n \propto n^{-1-2^{-p}}
\end{equation}

\section{Finite Size Effects}
\label{finite}

Epidemic outbreaks obviously involve finite populations. In other applications of branching processes the finiteness also plays a crucial role.  Finite size effects are especially pronounced for the critical branching processes \cite{rr,ML,bk}. Therefore it is important to understand how the basic characteristics such as the average outbreak size \cite{ML,bk} vary with population size $N$.

In a standard SIR model individuals of a population of size $N$ are either susceptible to the infection (S), infected (I), or recovered (R), who cannot get infected again. In a multistage model infected individuals can be at different stages of infection, and for all (except the last) stages recovery means passing to the next stage. The recovery rates are equal to one as in the infinite-population limit. The infection rate, however, is only initially equal to unity (as the process is critical), and it decreases with time since only susceptible individuals can get infected, and their number decreases. More precisely, the infection rate at any time is the ratio of the total number of susceptible individuals to the total population size.

The average size $\langle n\rangle=\sum_{n\geq 1} nA_n$ of the epidemic outbreak diverges in an infinite system due to the power-law tail \eqref{An_sol} of the probability distribution $A_n$. Setting the upper bound at the population size $N$ one estimates the average size of the epidemic outbreak 
\begin{equation}
\label{naive}
\langle n\rangle_{\text{naive}} \propto \sum_{n=1}^N n^{-1/2}\propto \sqrt{N}
\end{equation}
This heuristic argument is actually based on the tacit assumption that a finite fraction of the population [as reflected by the upper limit in the sum in \eqref{naive}] may get infected. It turns out that the actual average outbreak size is much smaller because the epidemic outbreak weakens as more individuals become infected. Let $M$ be the maximal outbreak size. The same estimate as in \eqref{naive} now gives $\langle n\rangle\propto \sqrt{M}$. On the other side, the effective infection rate is equal to $(N-M)/N=1-M/N$, and the average size of an outbreak in such sub-critical branching process is $N/M$, see \eqref{nav}. These two estimates should match, $\sqrt{M}\propto N/M$, from which $M\propto N^{2/3}$ thereby amending the naive estimate \eqref{naive} to a self-consistent estimate \cite{bk}
\begin{equation}
\label{real}
\langle n\rangle\propto \sqrt[3]{N}
\end{equation}
The scaling law \eqref{real} gives the correct asymptotic behavior. This interesting result was established by Martin-L\"{o}f \cite{ML}; several more recent studies \cite{bk,KS,ML2,Dutch,BK} confirm \eqref{real} and analyze other finite-size effects.  

Consider now the two-type CBP in the context of epidemic processes. We can think about $A$s as individuals carrying light infection which can be transmitted to susceptible individuals and evolve into deadly infection. The latter can be also be transmitted to susceptible individuals or it can cause death. We need to find the scaling behavior of the maximal outbreak sizes $M_A$ and $M_B$. Having established $M_A$ and $M_B$, one can use the power-law tails \eqref{An_sol} and \eqref{Bn_sol} to deduce the average outbreak sizes
\begin{equation}
\label{nAB:av}
\langle n\!_A\rangle\propto \sqrt{M_A}\,, \quad \langle n\!_B\rangle\propto M_B^{3/4}
\end{equation}
We again obtain an estimate $\sqrt{M_A}\propto N/M_B$. A naive conjecture is that the maximal outbreak sizes are comparable. In this situation 
\begin{equation}
\label{equal}
\begin{split}
&M_A\propto M_B\propto N^{2/3}\\
&\langle n\!_A\rangle\propto \sqrt[3]{N}\,, \quad \langle n\!_B\rangle\propto \sqrt{N}
\end{split}
\end{equation}

We know, however, that different types follow different scaling laws, specifically $A_n\propto n^{-3/2}$ and $B_n\propto n^{-5/4}$, and hence it seems much more plausible that $M_A\ll M_B$. In this case we use 
\begin{equation}
\label{MM}
\sqrt{M_A}\propto \frac{N}{M_B}\,, \quad \sqrt{M_A}\,\frac{N}{M_B}\propto M_B^{3/4}
\end{equation}
The second relation in \eqref{MM} is understood by noting that Eq.~\eqref{nav} with $\alpha=1-M_B/N$ gives $N/M_B$ for the average size of the outbreak caused by a single $B$, so in the case when the number of the seed individuals of type $B$ is equal to $\sqrt{M_A}$, we get $\sqrt{M_A}\,N/M_B$ for the average size of the outbreak. {}From \eqref{MM} we obtain the scaling of the maximal outbreak sizes
\begin{equation}
\label{non-equal:M}
M_A\propto N^{\frac{6}{11}}\,, \quad M_B\propto N^{\frac{8}{11}}
\end{equation}
which is combined with \eqref{nAB:av} to yield the average sizes
\begin{equation}
\label{non-equal}
\langle n\!_A\rangle\propto N^{\frac{3}{11}}\,, \quad 
\langle n\!_B\rangle\propto N^{\frac{6}{11}}
\end{equation}

\begin{figure}
\centering
\includegraphics[scale=0.66]{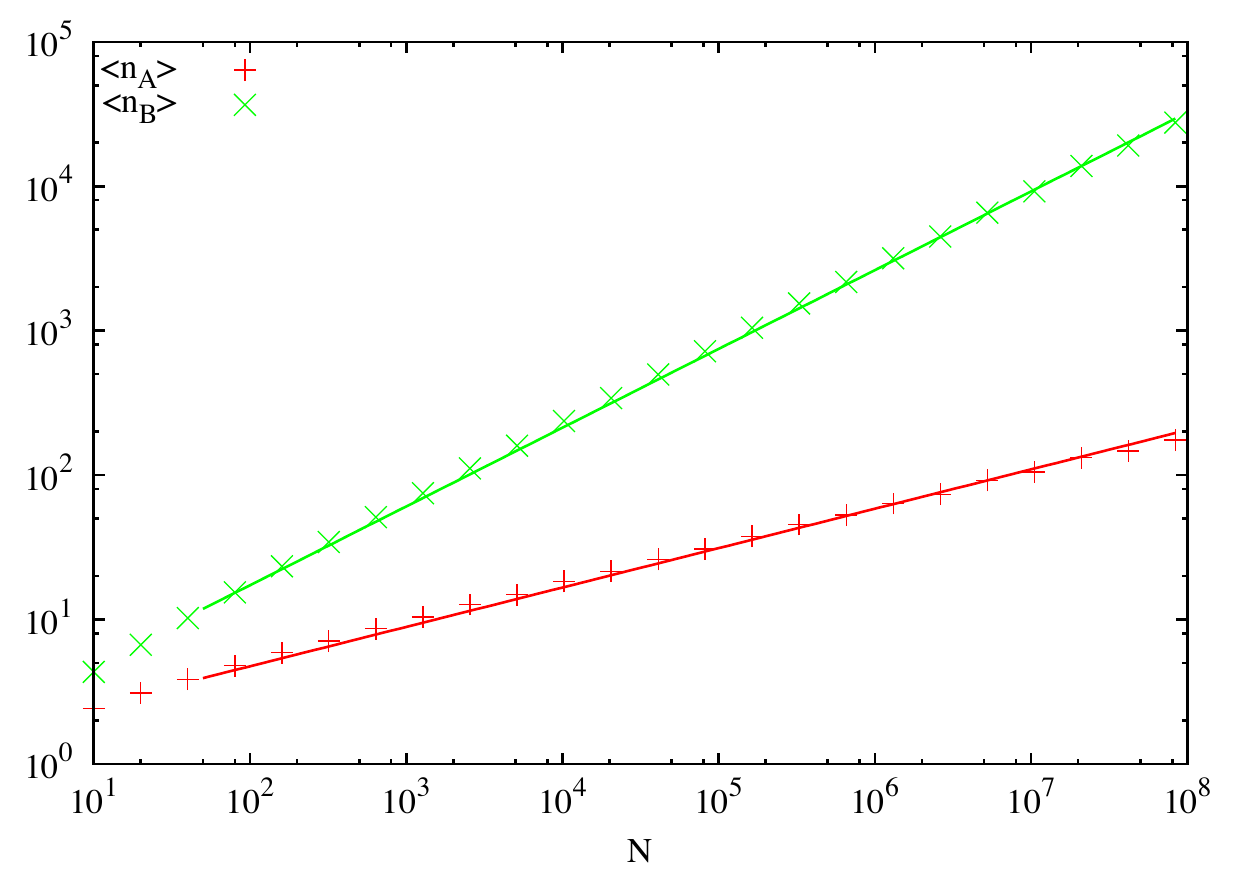}
\caption{Mean outbreak sizes arising in simulations for a two-type CBP in finite populations from an average over $10^5$ runs. The slopes of the lines are given by \eqref{non-equal}.}
\label{fig_s2}
\end{figure}

\begin{figure}
\centering
\includegraphics[scale=0.66]{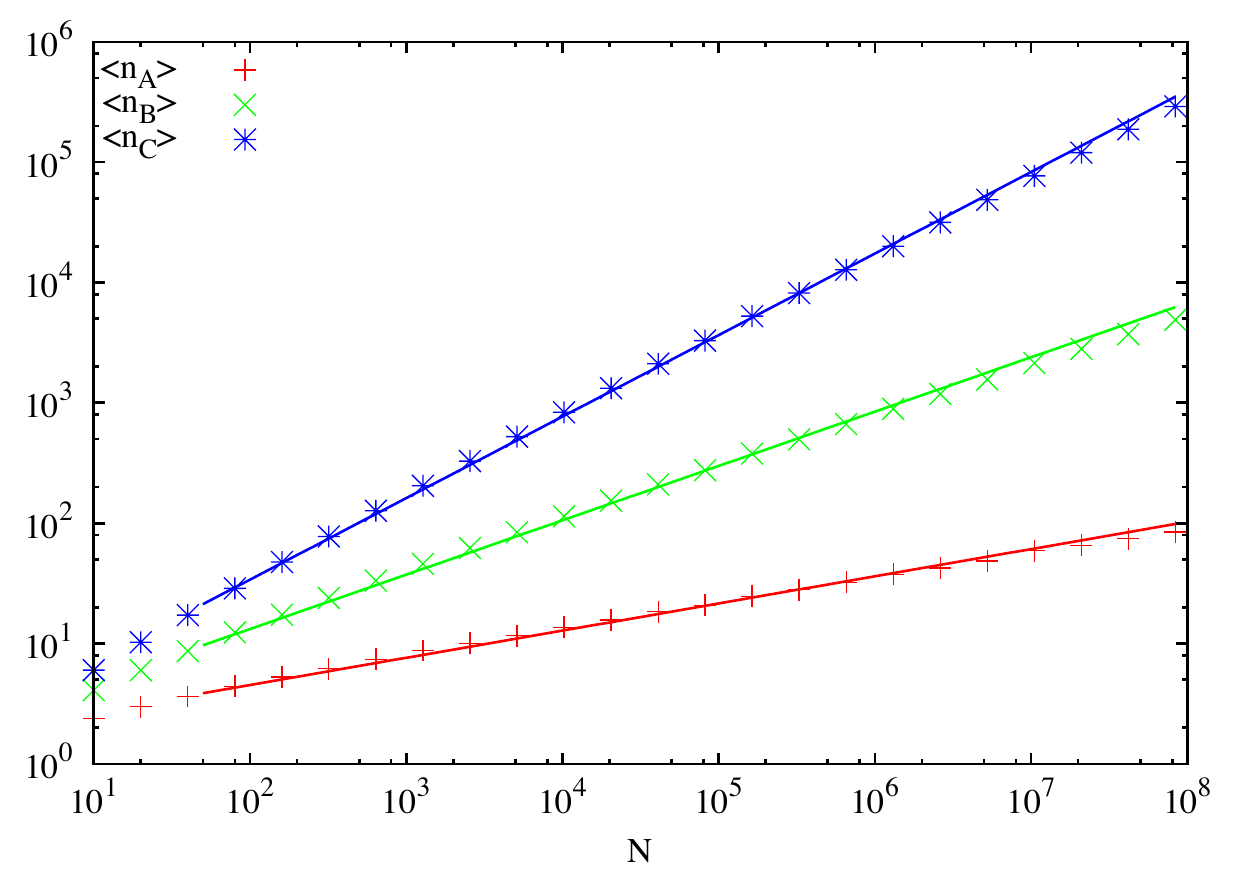}
\caption{Mean outbreak sizes arising in simulations for a three-type CBP in finite populations from an average over $10^5$ runs. The slopes of the lines are given by \eqref{n_ABC}.}
\label{fig_s3}
\end{figure}

For the three-type CBP the analog of \eqref{MM} reads
\begin{equation*}
\begin{split}
\frac{N}{M_C}                       &\propto M_A^{1/2}    \\ 
M_A^{1/2}\, \frac{N}{M_C} &\propto M_B^{3/4}\\
M_B^{3/4}\,\frac{N}{M_C}  &\propto M_C^{7/8}
\end{split}
\end{equation*}
{}from which the maximal outbreak sizes scale as 
\begin{equation}
\label{M_ABC}
M_A\propto N^{\frac{14}{31}}\,, \quad M_B\propto N^{\frac{56}{63}}\,, \quad M_C\propto N^{\frac{24}{31}}
\end{equation}
and the average outbreak sizes are given by 
\begin{equation}
\label{n_ABC}
\langle n\!_A\rangle \propto N^{\frac{7}{31}}\,, \quad \langle n\!_B\rangle\propto N^{\frac{14}{31}}\,, \quad
\langle n\!_C\rangle \propto N^{\frac{21}{31}}
\end{equation}

The predictions \eqref{non-equal} and \eqref{n_ABC} for the average outbreak sizes in the two- and three-type CBPs are in good agreement with simulations, see Figs.~\ref{fig_s2} and \ref{fig_s3}.

Generally for the $p-$type CBP one finds that the maximal and average outbreak sizes scale as 
\begin{equation}
\label{Mn}
M_q\propto N^{\lambda_p q/(1-2^{-q})}\,, \quad \langle n_q\rangle \propto N^{\lambda_p q}
\end{equation}
where $q=1,2,\ldots,p$ labels the types of individuals and 
\begin{equation*}
\lambda_p = \frac{p^{-1}(1-2^{-p})}{1+p^{-1}(1-2^{-p})}
\end{equation*}
The outbreaks of individuals of type $p$ (those which die rather than differentiate) are the largest:
\begin{equation*}
\begin{split}
&M_p                       \propto N^{1/[1+p^{-1}(1-2^{-p})]}\\
&\langle n_p\rangle \propto N^{(1-2^{-p})/[1+p^{-1}(1-2^{-p})]}
\end{split}
\end{equation*}

\section{Discussion}
\label{summary}

We investigated the multistage SIR infection process at the epidemic threshold. In an infinitely large population, we computed the outbreak size distributions and showed that these distributions have universal algebraic tails. Combining these infinite-population results and heuristic arguments we established scaling laws for finite populations. Specifically, we showed that the maximal and average outbreak sizes scale as powers of the population size. For instance, for the three-type critical branching process six non-trivial exponents, see Eqs.~\eqref{M_ABC}--\eqref{n_ABC}, describe the maximal and average outbreak sizes of the three types. We numerically verified the scaling laws and found an excellent agreement. 

A number of extensions are worth pursuing, especially those which would increase our understanding of the finite size scaling. Many of these extensions are easy to formulate, yet very challenging to analyze. For instance, one would like to prove the validity of the chief scaling results \eqref{Mn}. Even more demanding would be to determine the probabilities $A_n(N), B_n(N)$, etc. For the two-type CBP, for instance, the ratios of the probabilities $A_n(N), B_n(N)$ to their infinite-population values $A_n(\infty), B_n(\infty)$ given by Eqs.~\eqref{An_sol}, \eqref{Bn_sol}, are expected to exhibit scaling behaviors
\begin{equation}
\label{scaling}
\frac{A_n(N)}{A_n(\infty)} = F\!\left(\frac{n}{N^{6/11}}\right), \quad
\frac{B_n(N)}{B_n(\infty)} = G\!\left(\frac{n}{N^{8/11}}\right)
\end{equation}
The scaled sizes in Eq.~\eqref{scaling} are fixed by the scaling laws \eqref{non-equal:M} for the maximal outbreak sizes. 

It would be also interesting to determine the duration, both average and maximal, of the outbreaks; for the classical critical SIR process, this problem has been investigated in \cite{rr,bk,BK}. 

Finally, we stress that in this paper we limited ourselves to stochastic processes without any spatial structure. Needless to say, in applications the spatial or network structure of the infected domain play an important role \cite{PG,LG,Murray_II,Gunter,Ziff,kessler,odor}. The generalizations to such settings form an interesting avenue for future research.

\section*{Acknowlwdgements}

We are grateful to Burak Buke for discussions and suggestions. PLK acknowledges a support from NSF grant CCF-0829541.

\appendix
\section{Derivation of two-type probabilities}

We have obtained the explicit expression \eqref{2typeexp} by differentiating the generating function
\begin{equation}
\label{backdef}
 P_{m,n} = \frac{1}{m! n!} \partial_x^m \partial_y^n  \mathcal{P} |_{x=y=0}
\end{equation}
Here we provide some details. First, we differentiate the generating function \eqref{gen2type} with respect to $x$ to obtain
\begin{equation}
\label{diff_x}
 \partial_x^m \mathcal{P} = \frac{(2m-3)!!}{2^m} \frac{(1-\sqrt{1-y})^m}{[1-x(1-\sqrt{1-y})]^{m-1/2}}
\end{equation}
We can set $x=0$ on the right-hand side of \eqref{diff_x} since it won't affect $P_{m,n}$ defined by \eqref{backdef}. Hence we should differentiate $\partial_x^m \mathcal{P}|_{x=0} = \frac{(2m-3)!!}{2^m} (1-\sqrt{1-y})^m$ with respect to $y$ only. At each time we differentiate $(1-\sqrt{1-y})^m$, the number of terms grow by one until $n=m$, at which point the $m$ terms are proportional to $(1-\sqrt{1-y})^l$ with $l=0,\dots,m-1$. Differentiating further won't increase the number of terms.  Therefore
\begin{equation*}
\begin{split}
 &\partial_y^n (1-\sqrt{1-y})^m =\\
 &\sum_{k=\mathrm{max}(1,n-m+1)}^n \frac{s_{n,k}\, m!\, (1-\sqrt{1-y})^{m-n-1+k}}{2^n (m-n-1+k)!\, (1-y)^{(n+k-1)/2}}
\end{split}
\end{equation*}
This can be proved by induction and by using the recursion formula \eqref{recurs} for $s_{n,k}$. Setting $y=0$ in the sum on the right-hand side, we are left with only one term (the smallest $k$ term). Accounting for all factors in \eqref{backdef} we arrive at the announced result \eqref{2typeexp}.

\end{document}